\documentclass[twocolumn]{aastex631}

\usepackage{amsmath}
\usepackage{amssymb}
\usepackage{braket}
\usepackage{amssymb}
\usepackage{color}
\usepackage{xcolor}
\usepackage{graphicx}
\usepackage{latexsym}
\usepackage{listings}
\usepackage{mathrsfs}
\usepackage{letltxmacro}
\usepackage[normalem]{ulem}

\usepackage{subfigure}

\usepackage{verbatim}
\usepackage{tabularx}
\usepackage{ragged2e}
\usepackage{multirow}
\usepackage{amsbsy}

\usepackage{numprint}
\usepackage{makecell}
\usepackage[utf8]{inputenc}

\newcommand{\R}{\mathbb{R}}
\newcommand{\N}{\mathbb{N}} 

\newcommand{\V}{\mathcal{V}}

\newcommand{\F}{\mathcal{F}}

\newcommand{\M}{\mathcal{M}}

\newcommand{\I}{\mathcal{I}}

\newcommand{\B}{\text{B}}

\newcommand{\nubhlight}{$\nu\texttt{bhlight}$${}$}

\npdecimalsign{.}
\npthousandsep{,}

\newcolumntype{Y}{>{\RaggedRight\arraybackslash}X}

\graphicspath{{./figures/}}

\sloppypar

\received{}
\revised{}
\accepted{}
\submitjournal{ApJL}

\graphicspath{{./}{figures/}}

\begin{document}

\title{Nucleosynthesis in Outflows from Black Hole-Neutron Star Merger Disks With Full GR$\nu$RMHD}
\shorttitle{Nucleosynthesis from BHNS Disks}
\shortauthors{Curtis et al.}

\correspondingauthor{Sanjana Curtis, Jonah Miller}
\email{sanjanacurtis@uchicago.edu, jonahm@lanl.gov}

\author[0000-0002-3211-303X]{Sanjana Curtis}
\affiliation{Department of Astronomy and Astrophysics, University of Chicago, Chicago  IL 60637}
\affiliation{Center for Theoretical Astrophysics, Los Alamos National Lab, Los Alamos NM 87544}
\affiliation{Center for Nonlinear Studies, Los Alamos National Lab, Los Alamos NM 87544}

\author[0000-0001-6432-7860]{Jonah M. Miller}
\affiliation{CCS-2, Computational Physics and Methods, Los Alamos National Lab, Los Alamos NM 87544}
\affiliation{Center for Theoretical Astrophysics, Los Alamos National Lab, Los Alamos NM 87544}

\author[0000-0003-0191-2477]{Carla Fr\"ohlich}
\affiliation{Department of Physics, North Carolina State University, Raleigh NC 27695}

\author[0000-0002-4375-4369]{Trevor Sprouse}
\affiliation{T-2, Nuclear and Particle Physics, Astrophysics, and Cosmology, Los Alamos National Lab, Los Alamos NM 87544}
\affiliation{Center for Theoretical Astrophysics, Los Alamos National Lab, Los Alamos NM 87544}

\author[0000-0003-1707-7998]{Nicole Lloyd-Ronning}
\affiliation{CCS-2, Computational Physics and Methods, Los Alamos National Lab, Los Alamos NM 87544}
\affiliation{Center for Theoretical Astrophysics, Los Alamos National Lab, Los Alamos NM 87544}

\author[0000-0002-9950-9688]{Matthew Mumpower}
\affiliation{T-2, Nuclear and Particle Physics, Astrophysics, and Cosmology, Los Alamos National Lab, Los Alamos NM 87544}
\affiliation{Center for Theoretical Astrophysics, Los Alamos National Lab, Los Alamos NM 87544}

\begin{abstract}


Along with binary neutron star mergers, the in-spiral and merger of a black hole and a neutron star is
a predicted site of $r$-process nucleosynthesis and associated kilonovae. For the right mass ratio, very large amounts of neutron-rich material (relative to the dynamical ejecta) may become unbound from the post-merger accretion disk. We simulate a suite of four post-merger disks with three-dimensional general-relativistic magnetohydrodynamics with time-dependent Monte Carlo neutrino transport.
We find that within $10^4 GM_{\rm BH}/c^3$ ($\sim 200-500$\,ms), the outflows from these disks are very close to the threshold conditions for robust $r$-process nucleosynthesis. For these conditions, the detailed properties of the outflow determine whether a full $r$-process can or cannot occur, implying that a wide range of observable phenomena are possible. We show that on average the disk outflow lanthanide fraction is suppressed relative to the solar isotopic pattern. In combination with the dynamical ejecta, these outflows imply a kilonova with both blue and red components.

\end{abstract}


\section{Introduction}

The merger of a black hole - neutron star (BH-NS) binary is a potential multi-messenger event.
If the merger results in the tidal disruption of the NS, copious quantities of neutron-rich material undergoing rapid neutron-capture (or $r$-process) nucleosynthesis are ejected. The radioactive decay of freshly-synthesized heavy isotopes powers an electromagnetic transient called a kilonova. Very neutron-rich outflows 
are typically rich in lanthanides (140 $\leq A < 176$), which are highly opaque to blue light, and thus produce a red kilonova. Less neutron-rich outflows synthesize a smaller mass-fraction of lanthanides, producing an optical or blue kilonova instead. 



Whether the NS is tidally disrupted by the BH depends on the properties of the progenitor binary, such as the NS and BH masses, the BH spin, and the unknown NS equation of state (EOS). While gravitational waves from two BH-NS mergers have been confidently detected so far (GW200105 and GW200115), no kilonova counterpart has yet been found, which is not unexpected given the properties of the binary components \citep{LIGO2021}. Since many detections of BH-NS mergers are expected in the coming decades, modeling these systems with greater fidelity and predicting possible nucleosynthesis outcomes is of critical importance.

However, the composition of material ejected from BH-NS mergers remains uncertain. From theoretical modeling, the two main types of ejecta are the dynamical ejecta and post-merger disk outflows. A part of the disrupted NS material is ejected from the system due to tidal forces, referred to as the dynamical ejecta, while the rest settles into an accretion disk around the BH. The outflows from this post-merger accretion disk can constitute a significant fraction of the total merger ejecta 
 and represent the focus of our work. Although a number of studies have investigated such (or similar) outflows, these ejecta are not yet fully understood \citep{JustComprehensive2015, Fernandez2020, Fujibayashi2020PhRvD.101h3029F, Fujibayashi2020PhRvD.102l3014F, Fujibayashi2020, Fujibayashi2022arXiv, Wanajo2022}.

The details of the $r$-process nucleosynthesis depend on the electron fraction ($Y_e$) of the material, which is set by neutrino-matter interactions. The properties of the post-merger disk driving the outflow are set by the interplay of magnetohydrodynamical effects and gravity. Thus, to obtain an accurate prediction of ejecta properties and resulting nucleosynthesis,
we need general relativistic neutrino radiation magnetohydrodynamics (GR$\nu$RMHD) simulations.

There is a long history of treating post-merger outflows \citep{Ruffert1997DiskWind,PophamNDAF, Kohri2005ApJ} with various levels of approximation for angular momentum transport and neutrino physics, and the literature has recently exploded \citep{SiegelGW170917,Siegel2018ApJ, Nouri2018PhRvD, Christie2019MNRAS, Fernandez2019MNRAS, De_2021,Just2022MNRAS,Murguia-Berthier_2021,2022MNRAS.513.2689F,2022arXiv221205628N}. An exciting recent development is the self-consistent evolution of a binary NS merger from inspiral through the remnant phase, seconds after merger in \cite{2022PhRvD.106b3008H}. 
Recently, \citet{FoucartMerger2020} and \citet{2022arXiv221005670F} performed merger calculations with Monte Carlo neutrino transport, however, they did not follow the post-merger disk.
To-date, the only full-transport GR$\nu$RMHD simulations have been performed with our code, \nubhlight~ \citep{Miller2019code} in \cite{Miller2019} and \cite{Miller2020}.

Here, we present the first three-dimensional full transport
GR$\nu$RMHD
simulations of BH-accretion disk systems produced in BH-NS mergers. We study the conditions in four post-merger disks, spanning a range of BH and disk masses, and predict nucleosynthesis in the disk outflows.

\section{The Methods}
\label{sec:methods}

We use the publicly available code \nubhlight~\citep{Miller2019code}, which uses operator splitting to
couple GR$\nu$RMHD via finite volume methods with constrained transport
to Monte Carlo neutrino transport. 
We solve the equations of general relativistic ideal
magnetohydrodynamics, closed with the SFHo EOS, described in
\citet{SFHoEOS} and tabulated in \citet{stellarcollapsetables}.

Neutrinos can interact with matter via emission, absorption, or
scattering. 
For emission and absorption, 
we use the charged and
neutral current interactions as tabulated in \citet{fornax} and
summarized in \citet{BurrowsNeutrinos}. Neutrino scattering is implemented
as described in \citet{Miller2019code}.  

We use a radially logarithmic, quasi-spherical grid in horizon penetrating coordinates
with $N_r\times N_\theta \times N_\phi = 192\times 168\times 66$ grid
points with approximately $3\times 10^7$ Monte Carlo packets.  Although
our code is Eulerian, we track approximately $1.5\times 10^6$ Lagrangian fluid packets (``tracer particles'') of which
approximately $5\times 10^5$ become gravitationally unbound. Our tracer particles are initialized within the disk so that they uniformly sample disk material by volume. 

We run our simulations for approximately $10^4 G M_{BH}/c^3$, which allows us to observe the disk in a quasistationary turbulent state.
Depending on BH mass, this translates to a different amount of physical time, ranging from $\sim$ 200 - 500ms.

To ensure adequate resolution, we utilize the MRI quality factor defined in \citet{Sano2004} and the radiation quality factor defined in \citet{Miller2019}. We find both to be adequate for all disks at all times.
For more information on our code implementation and verification, see \citet{Miller2019code}.

\section{The models} 
\label{sec:models}


We simulate four accretion disks 
representing possible systems formed in the aftermath of BH-NS mergers. 
In the case of high component spins, 
a very large fraction of mass can end up outside the BH in either tidal or dynamical ejecta \citep{Kruger2020}. While the current population of BHs measured by LIGO-Virgo indicates a small projection of component spins onto the angular momentum axis, it is important to explore what is possible to interpret future observations. We therefore also include models with high component spins, resulting in large disk masses.
The properties of our initial BH-disk configurations are listed in Table \ref{tab:setup}. Note that the disks vary not just in disk mass but with respect to several initial properties, such as BH mass and spin. 

\begin{table}
        \begin{center}
                \caption{Initial conditions
                \label{tab:setup}
        }

               \begin{tabular}{ccccccc}
                        \tableline \tableline
                        Label & $M_{BH}$ & $a$ & $M_d$ & $R_{in}$ & $s$ & $Y_e$ \\
                        (-) & $M_{\odot}$ & (-) & $M_{\odot}$ & km &  $k_b$/baryon & (-)  \\
                        \tableline
                        Disk 1 & 10   & 0.8 & 0.082   & 50     & 8 & 0.2 \\
                        Disk 2 & 6  & 0.75 & 0.1425 & 27    & 8 &  0.15 \\
                        Disk 3 & 7  & 0.9 &  0.25 & 40   &  4   &   0.1  \\
                        Disk 4  & 5  & 0.9 & 0.42 & 30 &  4   & 0.1 \\ 
                        \tableline
                \end{tabular}
        \end{center}
    \tablecomments{Disk label and corresponding remnant BH mass and spin, disk mass, inner radius of the disk, initial entropy and initial electron fraction. Disk masses are estimated based on the fitting formulae derived in \cite{Foucart2012}, assuming binary properties that favor NS disruption with the goal of producing a wide range of disk masses. The entropy and $Y_e$ for each disk were chosen based on the typical values obtained in dynamical BH-NS merger simulations (Francois Foucart, private communication).
    }
\end{table}

To form the accretion disk, we begin with a constant entropy and constant $Y_e$ torus in hydrostatic equilibrium around a Kerr BH, as described by \citet{FishboneMoncrief}. Our torus starts with a single poloidal magnetic
field loop with a minimum ratio of gas to magnetic pressure $\beta$ of 100. As
the system evolves, the magneto-rotational instability
\citep[MRI,][]{BalbusHawley91} self-consistently drives the disk to a
turbulent state, which provides the turbulent viscosity necessary for
the disk to accrete.

\begin{figure*}
    \begin{tabular}{cc}
    \includegraphics[width=0.48\textwidth]{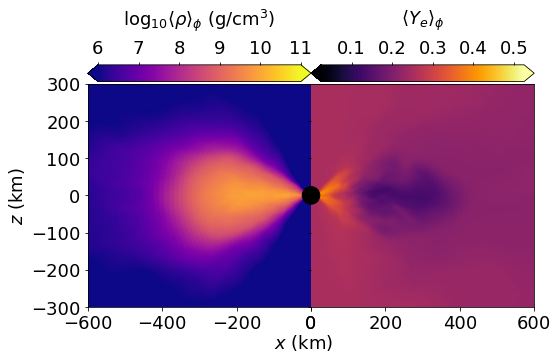}
    \includegraphics[width=0.41\textwidth]{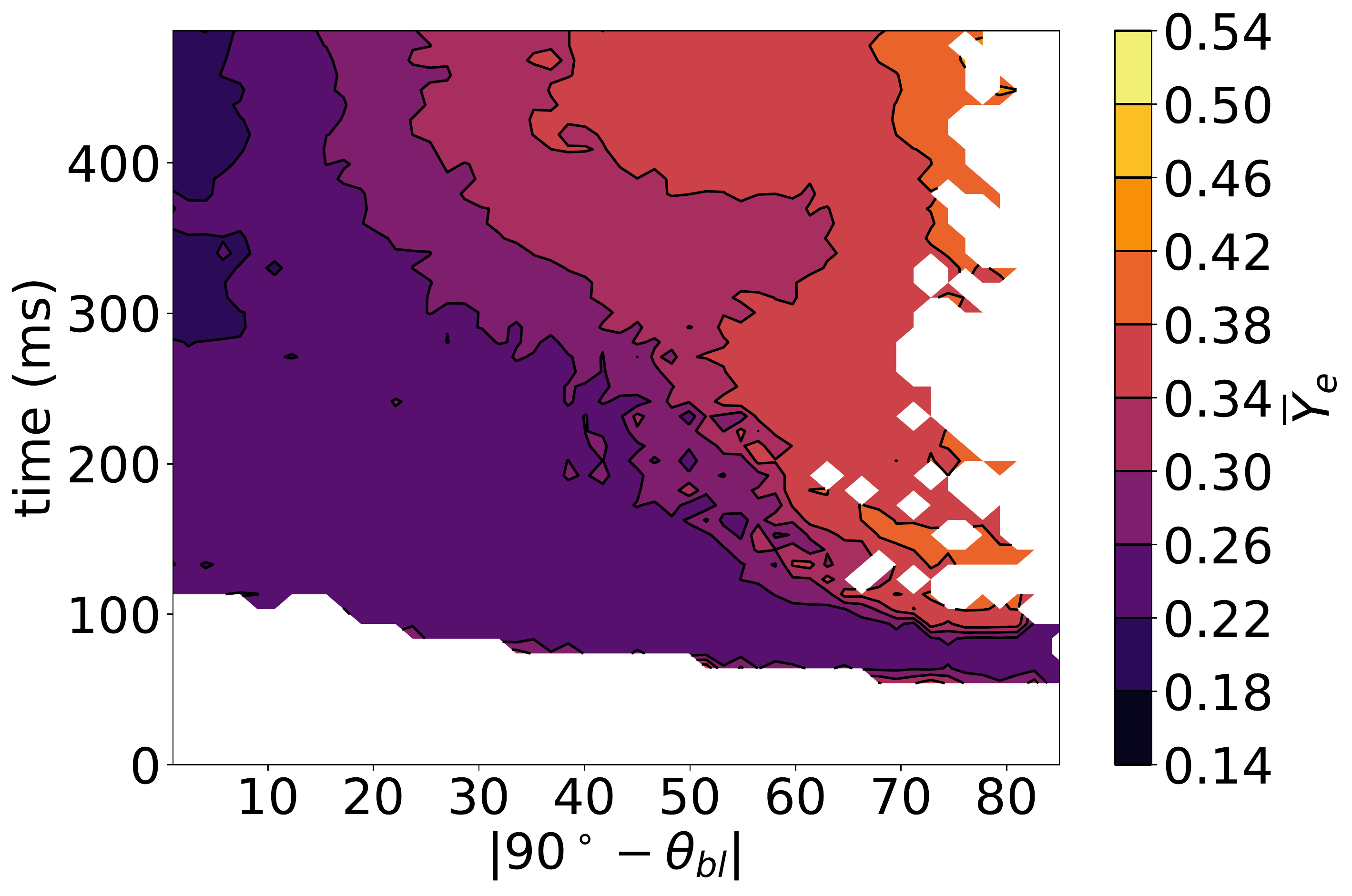}\\
    \includegraphics[width=0.48\textwidth]{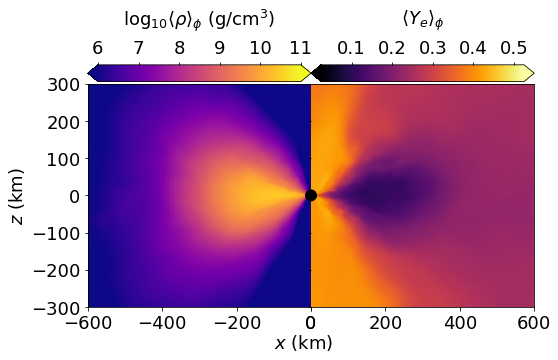}
    \includegraphics[width=0.41\textwidth]{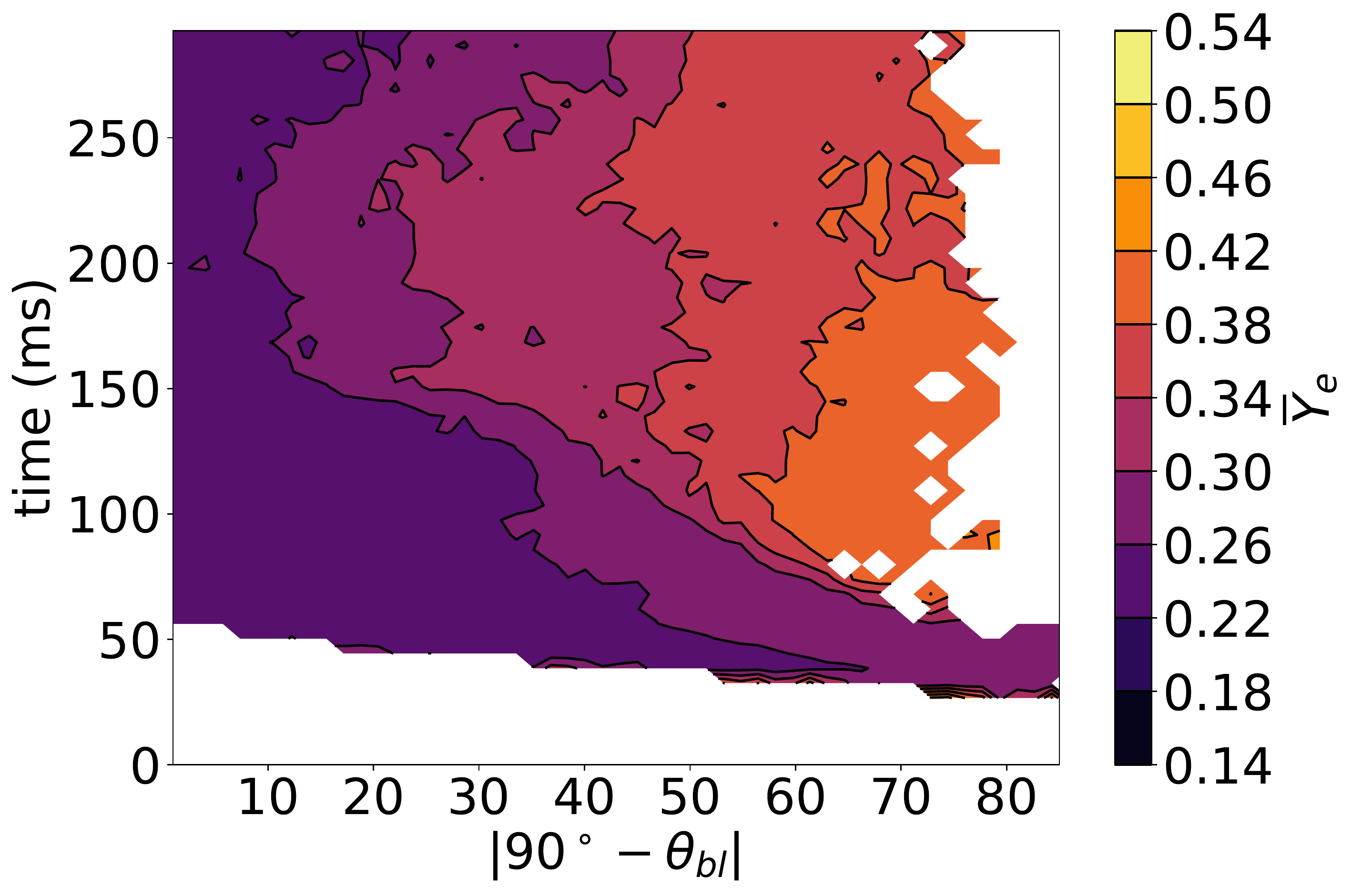}\\
    \includegraphics[width=0.48\textwidth]{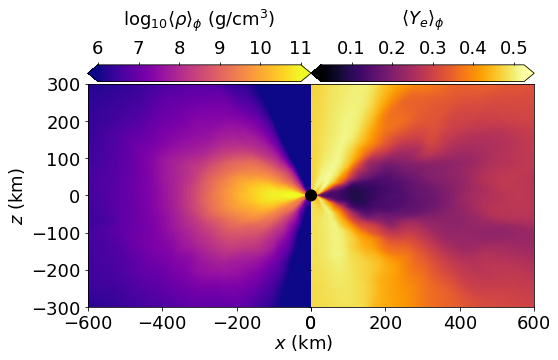}
    \includegraphics[width=0.41\textwidth]{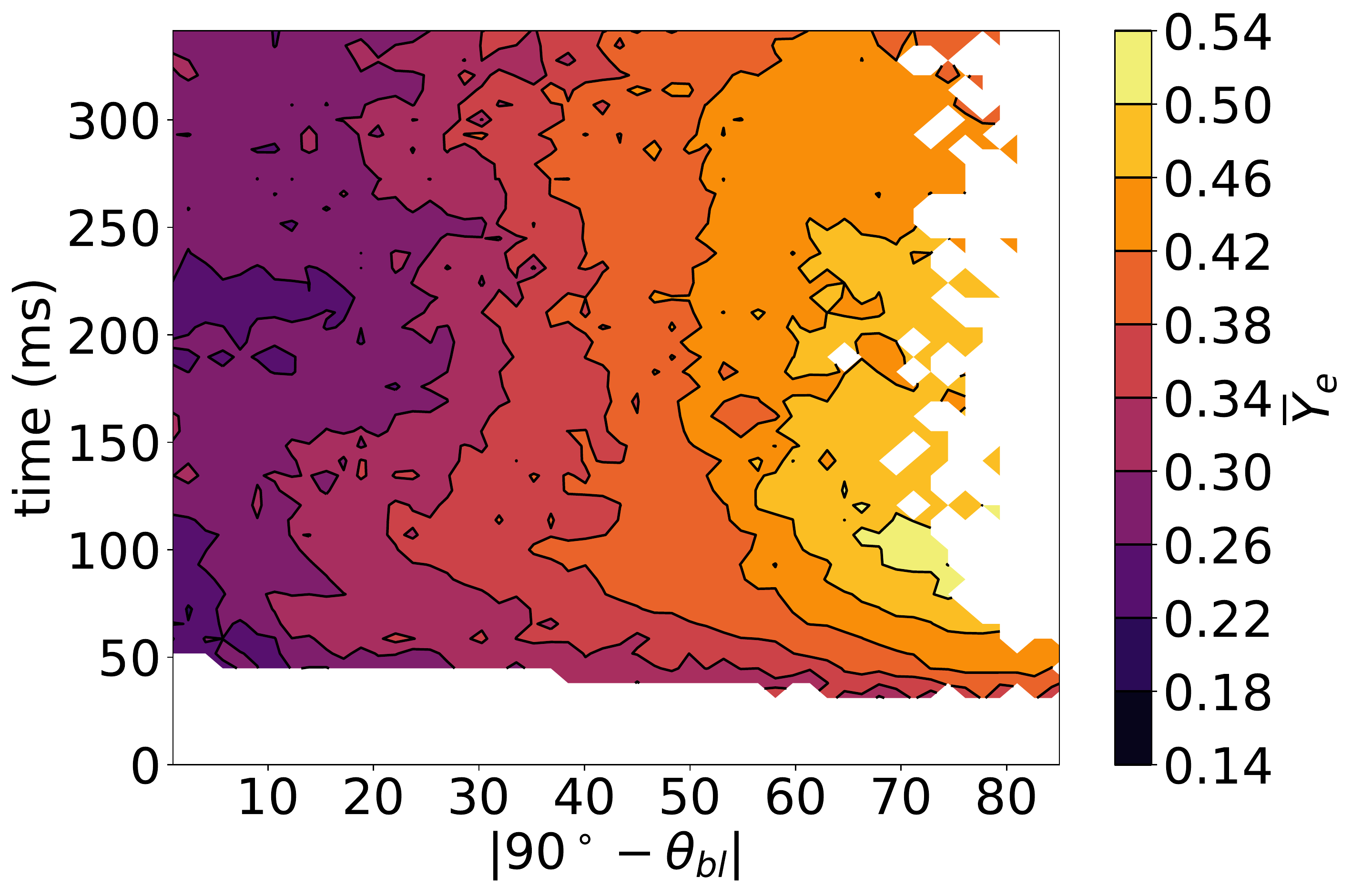}\\
    \includegraphics[width=0.48\textwidth]{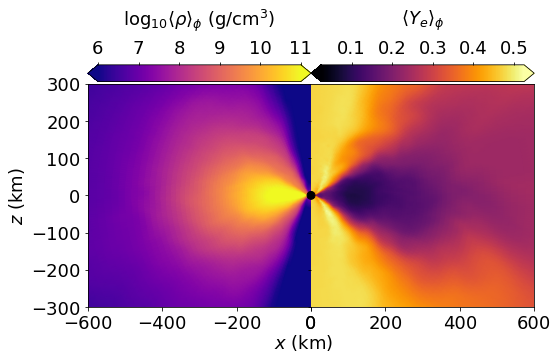}
    \includegraphics[width=0.41\textwidth]{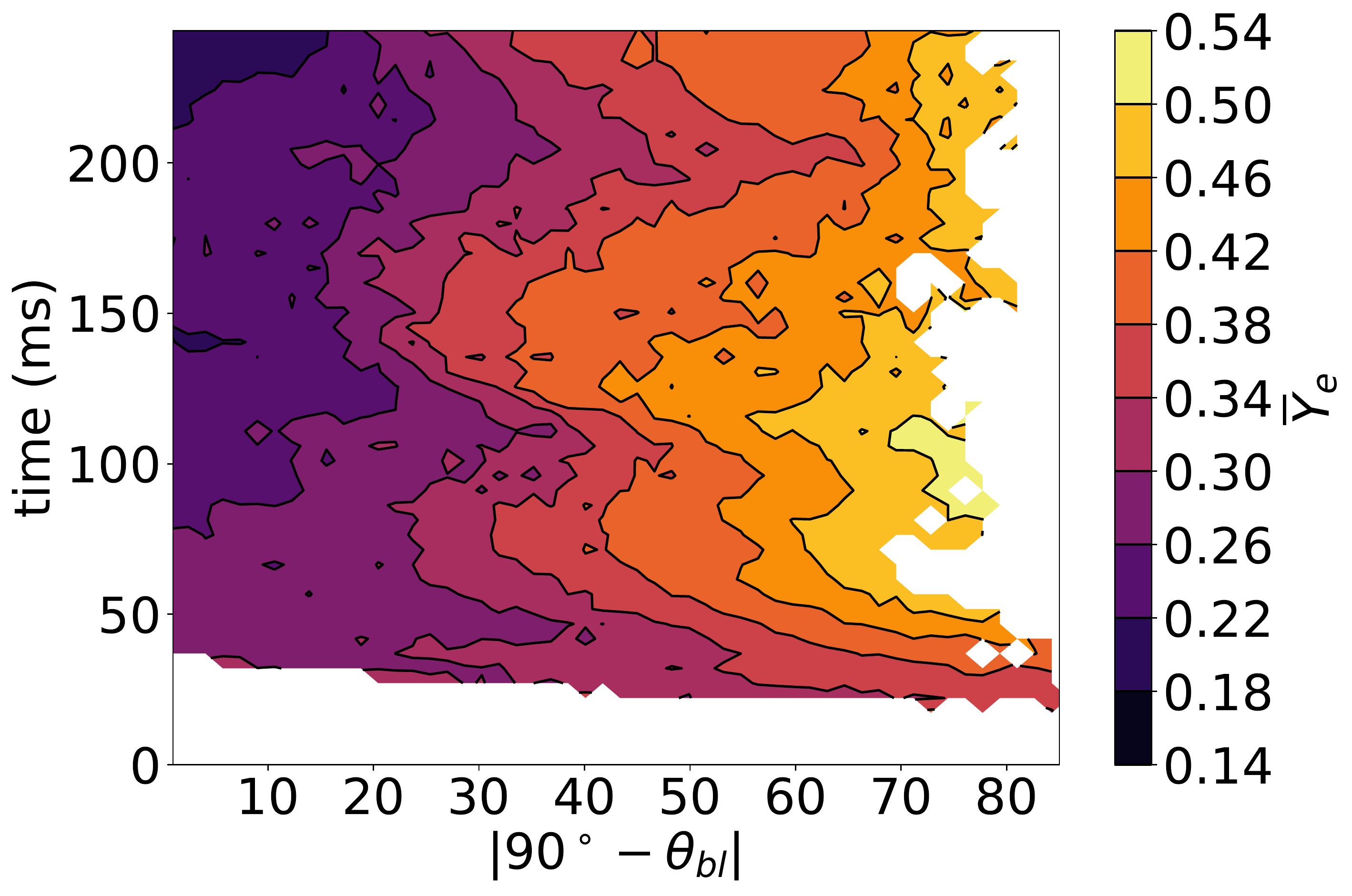}
    \end{tabular}
        \caption{Left: Azimuthally averaged density (left half) and $Y_e$ (right half) of each of the four disk models going from the least massive disk (top) to the most massive disk (bottom). The snapshots are taken at approximately 50ms. Right: Average $Y_e$ of gravitationally unbound material as a function of latitude and time. The angle off the equator is given by $|$90$^{\circ}$-$\theta_{bl}$$|$, where $\theta_{bl}$ is the Boyer-Lindquist angle. Note the different end times for the four simulations.} 
        \label{fig:disks_rho_ye}
\end{figure*}

\section{Results}
\label{sec:results}

\subsection{Disk Outflows}
\label{sec:disk:outflows}

As the disk accretes, turbulent viscosity transfers angular momentum outward, carrying with it material. Some of this material will become unbound - the viscous outflow - due to energy injection by turbulent energy dissipation and nuclear recombination, without neutrino cooling. In-fall of material transforms gravitational potential energy into thermal, kinetic, and electromagnetic energy, which powers thermally- and magnetically-driven winds off the ``surface'' of the disk and through the corona. A magnetically powered jet clears the polar region and entrains some material, carrying it out the poles.

Our ejecta are comprised of material with Bernoulli parameter $B_e>0$ at a radius of roughly 250 gravitational radii. Most of the ejected material is moving at mildly relativistic speeds $\lesssim$0.1--0.2 $c$. Due to computational cost, we did not run our simulations long enough to determine the total unbound mass from the disks. 
Thus, the ejecta masses obtained in the simulations are lower limits. 

Figure \ref{fig:disks_rho_ye} shows snapshots of the density and $Y_e$ of the four disks (left panels) and the evolution of the average $Y_e$ of the ejecta as a function of latitude and time (right panels). Typically, at early times, electron neutrinos get emitted near the equator and absorbed at relatively higher latitudes, driving the $Y_e$ in this region towards higher values. This imparts an angle-dependent structure to the $Y_e$, with $Y_e$ values increasing with the angle off the equator, and this structure persists over time. 
At late times, the optical depths are lower, the neutrinos are basically free-streaming, and the $Y_e$ evolution is dominated by neutrino emission instead. 

The properties of the disk outflow -- primarily $Y_e$, specific entropy $s$, and dynamical timescale $\tau_{\textrm{dyn}}$ -- set the $r$-process nucleosynthesis yields. 
Depending on $s$ and $\tau_{\textrm{dyn}}$, ejecta are expected to be lanthanide-free for $Y_e \gtrsim 0.22 - 0.3$ \citep{Lippuner2015}.
We record the $Y_e$ and $s$ of our ejecta, and compute $\tau_{\textrm{dyn}}$ as follows:
\begin{equation}
    \label{eq:tau:dyn}
    \tau_{dyn} = \frac{\int \frac{\rho}{\dot{\rho}} dt}{\int dt}
\end{equation}
where we take the integral over the time it takes for the temperature of the material to drop from approximately 10GK to 1GK.

\begin{figure}
\begin{tabular}{c} 
\includegraphics[width=0.45\textwidth]{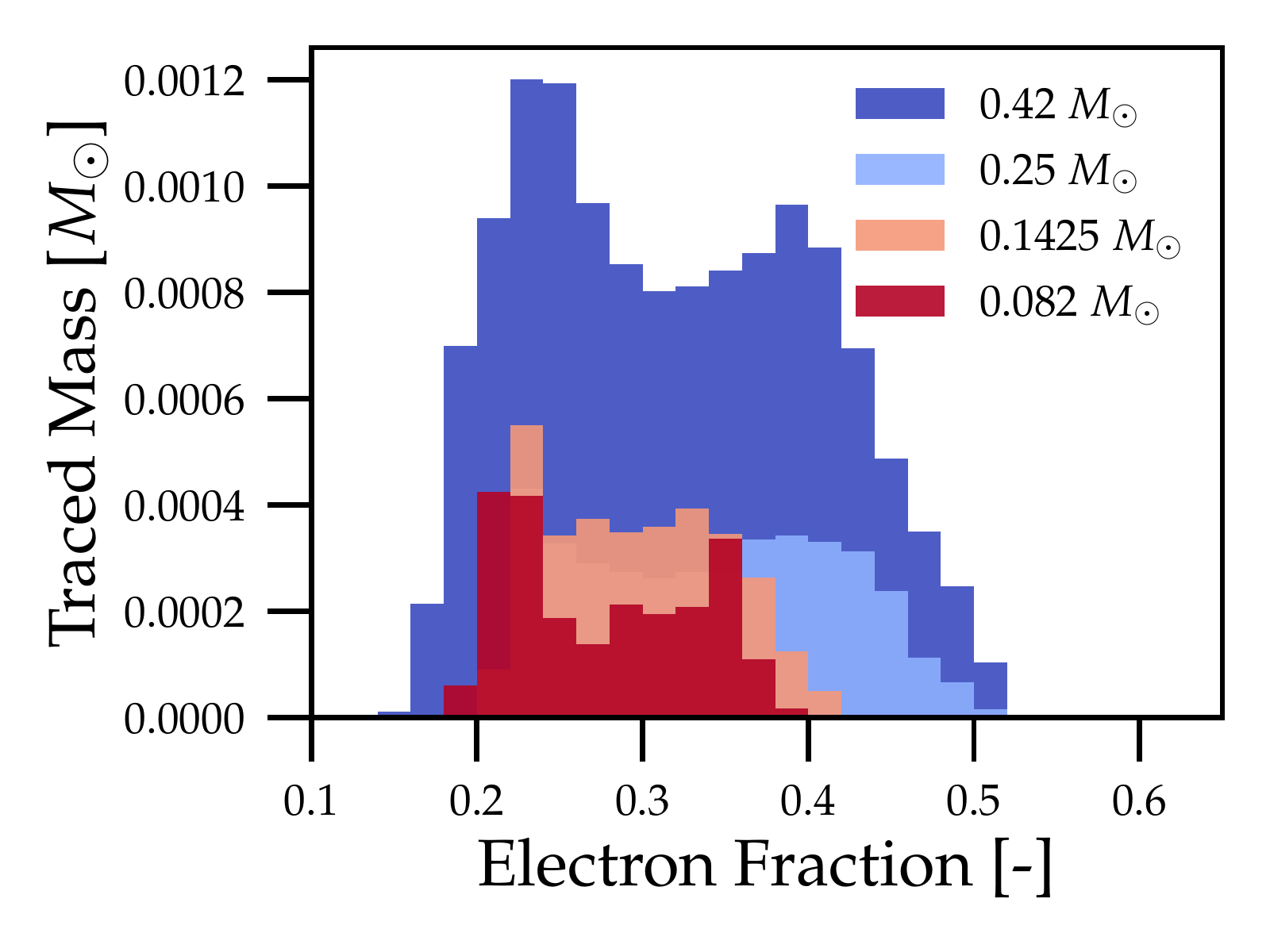}\\
\includegraphics[width=0.45\textwidth]{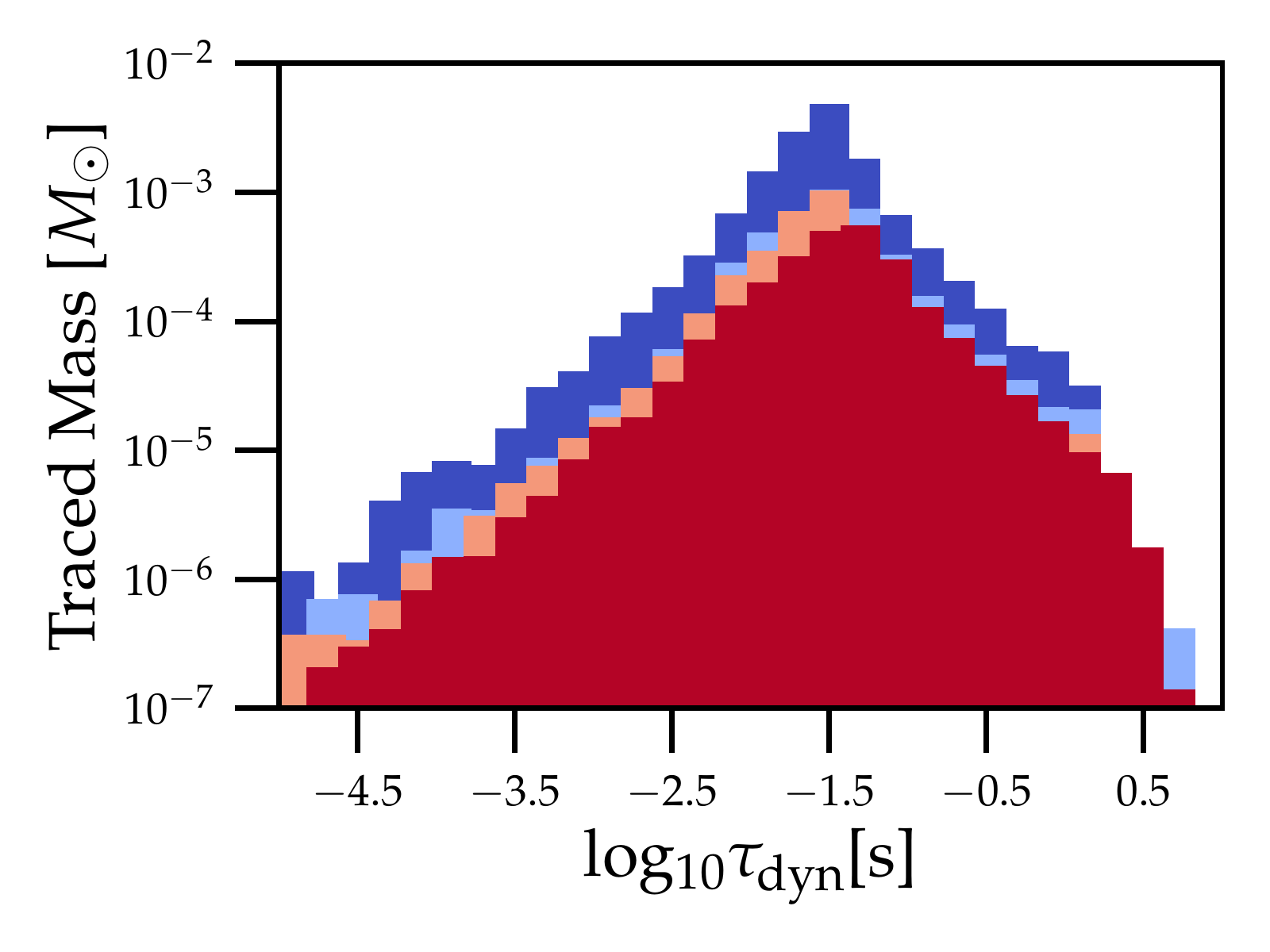}\\
\includegraphics[width=0.45\textwidth]{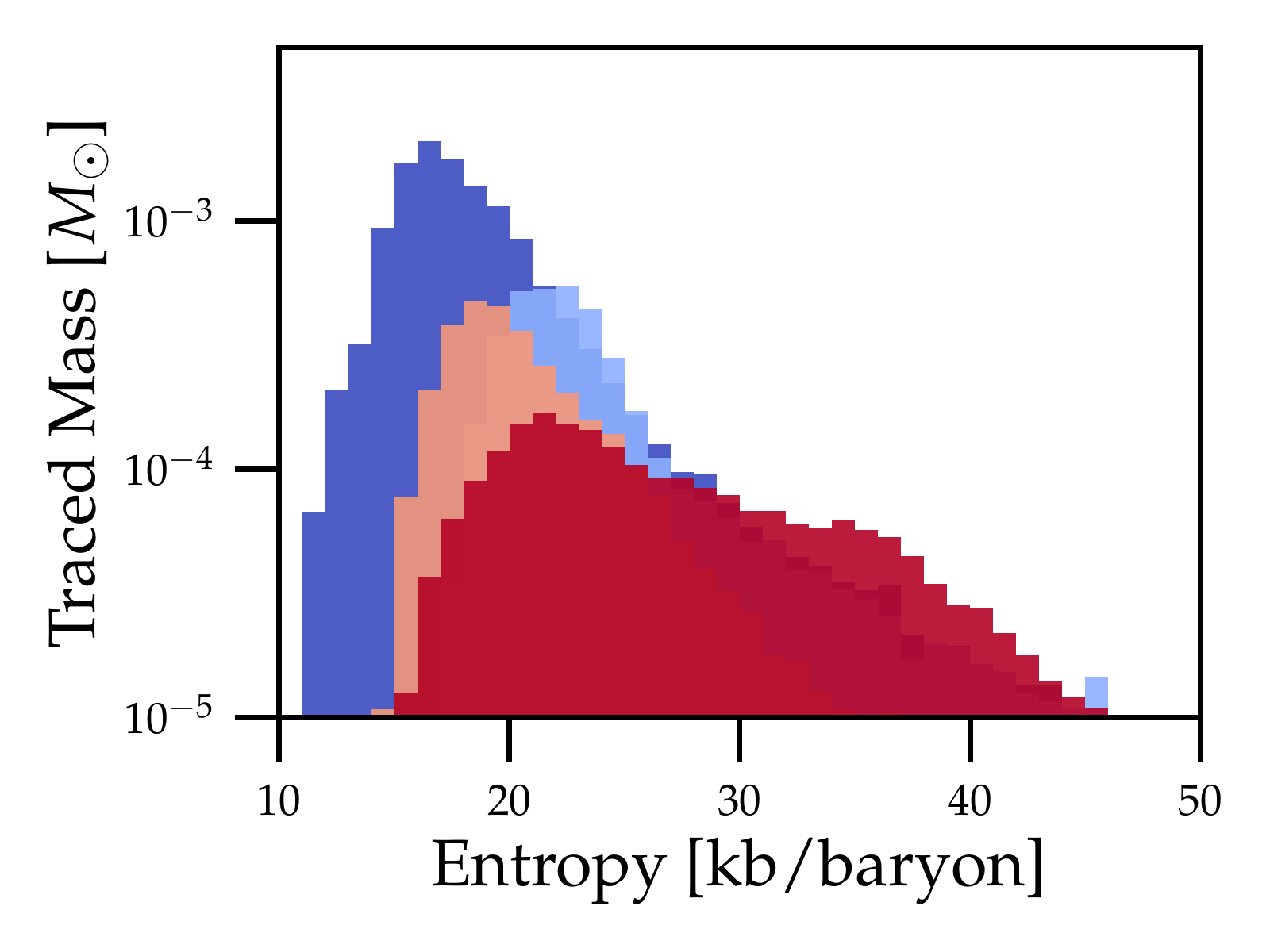}
\end{tabular}
        \caption{From top to bottom: Histograms representing the electron fraction, mean dynamical timescale, and entropy of the ejecta for all four BH-disk systems.}
        \label{fig:histograms}
\end{figure}

In Figure \ref{fig:histograms}, we present the $Y_e$, $\tau_{\textrm{dyn}}$, and $s$ of the total ejecta, for all four disks. We can see that the $Y_e$ of virtually all of the ejected material across the four disks lies above $\sim$0.2 and the distribution peaks at values between $\sim$ 0.2 -- 0.4. The $Y_e$ distributions have a double peaked structure, with the peak at lower $Y_e$ corresponding to neutron-rich ejecta present in the midplane and the peak at higher $Y_e$ corresponding to mass ejection at higher latitudes, where $Y_e$ is driven up by neutrino absorption. We note a broadening of the distribution towards higher maximum $Y_e$ with increasing disk mass.  

While the $Y_e$ varies meaningfully from one disk to another, $s$ and $\tau_{\textrm{dyn}}$ show little variation from disk to disk. The dynamical time is roughly 30ms for all four disks, with long tails. 
Most of the ejected material has entropies between 10--30~$k_B$/baryon, with only the tails extending to different values for the different disks, with the exception of Disk 1, which has a significant high angle--high entropy outflow component.

With respect to nucleosynthesis, the $Y_e$ distribution in these outflows is both broad and marginal. For entropies and dynamical timescales typical for these simulations, $Y_e \sim 0.24$ is the lanthanide turn-off point i.e. above this value, the lanthanide mass-fraction drops below $10^{-3}$ \citep{Lippuner2015}. These BH-NS disks have a substantial fraction of material close to this marginal value, consistent with a mildly suppressed $r$-process. 

\subsection{Nucleosynthesis}

To compute $r$-process yields, the tracer particles representing the gravitationally unbound material are post-processed using \texttt{PRISM} \citep{Mump2017, Sprouse2021}. The nucleosynthesis calculation for a tracer starts when its temperature falls below $\sim$10 GK and the initial composition is assumed to be the nuclear statistical equilbrium composition. The calculations are continued until 1 Gyr, assuming homologous expansion beyond the end time of a tracer.  

Nuclear data is implemented in our calculations following the prescription described in \cite{Zhu2018}. All relevant nuclear reactions, such as charged particle reactions~\citep{Cyburt2010}, neutron capture~\citep{Kawano2016}, photo-dissociation, $\beta$-decay~\citep{MumpQRPA+HF, Mumpower2018,MOLLER20191}, and fission are included. We supplement the datasets with the nuclear decay properties of the Nubase 2016 evaluation~\citep{NUBASE2016} and AME2016~\citep{AME2016} where appropriate.

\begin{figure}
\begin{center}
        \includegraphics[width=\columnwidth]{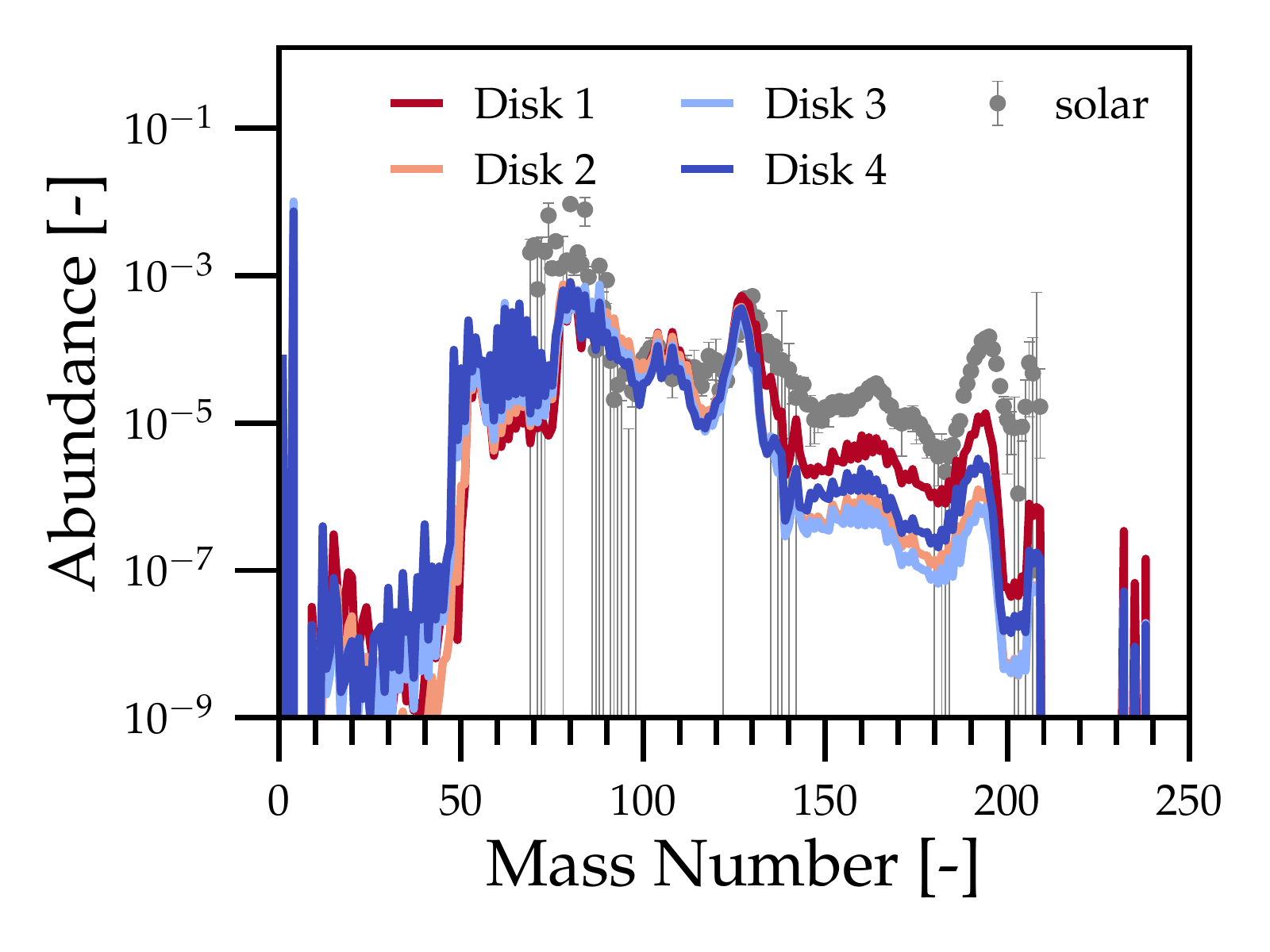}
        \end{center}
        \caption{Mass-weighted abundance as a function of mass number for outflows from all four accretion disks. Solar abundances from \cite{Arnould2007} are shown in gray and scaled to approximately match Disk 1 around the second peak.}
        \label{fig:nucleosynthesis}
\end{figure}

The final abundances computed with \texttt{PRISM} for all four disks are shown in Figure \ref{fig:nucleosynthesis}, showing a full range of $r$-process nuclei. 
However, the abundances of isotopes beyond the second $r$-process peak $A\sim130$ are suppressed to various extents relative to their solar values. Outflows from Disk 1 have the highest mass-fraction of elements beyond the second peak, including the lanthanides, while abundances of these isotopes are most suppressed for Disk 3, almost two orders of magnitude lower than solar values. Disk 4, which is the most massive disk, sits between these two abundance patterns. Abundances of Disk 2 outflows are very similar to those for Disk 3, with slightly higher values around the third peak. For Disks 3 and 4, we find enhanced production of nuclei with A $\sim$ 60-70 relative to Disks 1 and 2, associated mainly with the production of neutron-rich isotopes of Nickel and Zinc in the relatively high $Y_e$ ejecta component ($\gtrsim$0.4) present for these two disks. This suggests that it is possible to eject Zinc along with $r$-process elements for some fraction of BHNS mergers. Such an abundance signature is thus not unique to magnetorotational supernovae, which have been invoked to explain high [Zn/Fe] ratios seen in metal-poor stars \citep{Nishimura2017, Tsujimoto2018}.


These abundance patterns are consistent with our expectations based on the $Y_e$, $s$ and $\tau$ distribution of the ejecta from the disks. We compute the lanthanide fractions in the ejecta, defined as the mass of lanthanides divided by the mass of all neutron-capture elements, and estimate the lanthanide fraction for the total ejecta (dynamical and disk) in Table \ref{tab:summary}.
 
\subsection{Late Time}
\label{subsec:latetime}

Accretion, and thus outflow, has not finished during our ~\nubhlight~ simulations. By the final simulation time, 2.8\%, 2.2\%, 1.6\% and 3.1\% of the original disk mass has become gravitationally unbound for Disks 1--4, respectively. The fraction of the original disk mass that has accreted during the simulation is 37.5\%,
27.7\%, 41.5\%, and 37.9\%
respectively. Dividing the unbound mass by the accreted mass, we find that the percentage of the accreted disk mass that has become unbound during the course of the simulation is 7.5\%, 8.3\%, 3.7\% and 8.3\% for Disks 1--4, respectively.
We thus estimate that after a disk has fully accreted, a total of roughly 8\% of the disk mass will become gravitationally unbound. However, 
other studies suggest values as high as  40\% \citep{SiegelGW170917,Fernandez2019MNRAS,Christie2019MNRAS}. We provide the estimated disk ejecta mass range in Table \ref{tab:summary}.

As the disk drains and density falls, neutrino absorption becomes subdominant to emission,
and eventually the electron fraction in the disk will be driven towards its value at emission equilibrium (where the rate of electron fraction increase from the emission of electron anti-neutrinos is balanced exactly by the rate of decrease from the emission of electron neutrinos).
As a lower bound on the electron fraction of the outflow that might become unbound after the simulation time ends, we report the spherically averaged, density-weighted emission equilibrium $Y_e$ \citep{Miller2020},
$$\left\langle Y_e\right\rangle_{\text{sadw}}^{\text{em}} = \frac{\int d^3 x\sqrt{g} \rho Y^{\text{em}}_e}{\int d^3 x \sqrt{g}\rho},$$
where $Y_e^{\text{em}}$ is computed by solving for the $Y_e$ at a given density and temperature that equilibrates the emission rates of electron neutrinos and their antiparticles.

We find that $\left\langle Y_e\right\rangle^{\text{em}}$ is 0.39, 0.28, 0.24 and 0.21 for Disk 1, Disk 2, Disk 3 and Disk 4 respectively.
This estimate indicates that perhaps the most massive disks \textit{may} produce lanthanide-rich outflow at later times. However, we emphasize that the emission equilibrium value is a lower bound, and for the most massive disks in particular, absorption will matter longer due to high neutrino opacities, and likely raise $Y_e$ in the outflow.

\begin{table*}
        \begin{center}
                \caption{Ejecta masses ($M_{\odot}$) and lanthanide fractions (-) 
                \label{tab:summary}
        }
                \begin{tabular}{cccc|c|c|c}
                        \tableline \tableline
                        Label & Disk Mass & Disk Ejecta & $X_{\mathrm{La, disk}}$ & Estimated Dynamical Ejecta & Estimated Disk Ejecta & $X_{\mathrm{La, total}}$ \\
                        \tableline
                        Disk 1 & 0.082 & 0.0023 & 0.02  &  
                        0.04--0.05 & 0.0066 -- 
                        0.0328
                        & 0.03 -- 0.04 \\
                        Disk 2 & 0.1425 & 0.0032 & 0.0035    & 
                        0.02--0.04 & 0.0114 -- 
                        0.057
                        & 
                        0.013 -- 0.03\\
                        Disk 3 & 0.25 & 0.0039  &  0.0031    & 
                        0.04--0.06 & 0.02 -- 
                        0.1 &  
                        0.014 -- 0.03 \\
                        Disk 4 & 0.42 & 0.013 & 0.0077  & 
                        0.07--0.08 & 0.0336 -- 
                        0.168
                        & 
                        0.017 -- 0.03 \\
                        \tableline
                \end{tabular}
        \end{center}
    \tablecomments{Ejecta component masses and lanthanide fractions for all four disks, both computed from \nubhlight~ simulations (columns 3 and 4) and estimated. Dynamical ejecta mass estimates (column 5) were provided by Francois Foucart (private communication) based on \cite{Kawaguchi2016} and \cite{Kruger2020}. The lanthanide fraction of the dynamical ejecta is assumed to be solar,  $X_{\textrm{La}}$$\sim$~0.04, using solar abundances from \cite{Arnould2007} \citep{Ji2019}. Total disk ejecta masses (column 6) are estimated to be 8--40\% of the disk mass and the lanthanide fraction is assumed to be the same as that of the disk ejecta produced during the course of the simulation. $X_{\textrm{La, total}}$ (column 7) combines the total dynamical and disk ejecta estimates with their corresponding lanthanide fractions to estimate an overall lanthanide fraction for the total BH-NS ejecta.}

\end{table*}

\section{Summary and Discussion}

In this paper, we present the first full transport GR$\nu$RMHD simulations of post-merger accretion disks resulting from BH-NS mergers. We compute nucleosynthesis yields in the disk outflows for four different BH-disk configurations. We find copious production of $r$-process elements in these ejecta, however, the production of elements beyond the second peak is suppressed by up to a factor of $\gtrsim$10--100 relative to solar abundances. This is consistent with the ejecta $Y_e$ being driven to values above the threshold for lanthanide production due to neutrino-matter interactions. 

In Table \ref{tab:summary}, we present the total ejecta mass produced during the course of \nubhlight~simulations along with the lanthanide fraction $X_{\textrm{La}}$ in these ejecta. The lanthanide fraction distribution of metal-poor stars, which peaks at log $X_{\textrm{La}} \sim -1.8$, provides an observational test for the composition of merger ejecta.  
While Disk 1 produces $X_{\textrm{La}} \sim 10^{-2}$, Disks 2, 3, and 4 produce $X_{\textrm{La}} \sim$ a few times 10$^{-3}$. Disk outflows from the latter three disks thus represent an interesting ejecta component distinct in their nucleosynthetic signature from dynamical ejecta, which are expected to be very neutron-rich and typically have lanthanide fractions of the order of 10$^{-2}$. Combining the early disk ejecta with estimates of the late time disk ejecta and the dynamical ejecta gives us a sense of the overall nucleosynthesis in such mergers. While $X_{\textrm{La}}$ of ejecta from disks alone lies below the peak of the distribution observed in metal-poor stars, when tied with the dynamical ejecta, BH-NS mergers can produce lanthanide fractions close to that of the typical metal-poor star.

Given the sophisticated microphysics and neutrino transport in $\nu$bhlight, we are uniquely poised to comment on the nature of BH-NS kilonovae. Typically, for $X_{\textrm{La}} \gtrsim 3\times 10^{-3}$, the kilonova peaks in the near-infrared J band around 1 day \citep{Even2020}. However, the total $X_{\textrm{La}}$ is less relevant here compared to how ejecta composition varies with latitude as well as its overall morphology \citep{Korobkin2021}. Since the $Y_e$ of our outflows varies significantly with angle off of the midplane, the observed character of the kilonova will depend heavily on viewing angle. In general, the polar ejecta have higher abundances of isotopes below $A\sim120$ and lower abundances of isotopes at/between the second and third peaks, relative to the equatorial ejecta. This will shift the kilonova blue-ward due to a decrease in the lanthanide and actinide abundances.  
Thus, an early blue wind-produced kilonova may be visible if the remnant is viewed close to the polar axis. The observation of such a kilonova from a BH-NS merger will be a particularly exciting discovery since, unlike binary NS mergers, any optical kilonova component can only be produced by the post-merger disk outflows. For BH-NS mergers, the dynamical ejecta lie on the equatorial plane (due to their tidal origin) and are very neutron rich, yielding a lanthanide-rich contribution to the kilonova \citep{Shibata2019ARNPS}. 


Several long gamma ray bursts---namely GRB 060614 \citep{DellaValle2006,GalYam2006,Zhang2007}, GRB 211221A \citep{Rastinejad2022,Xiao2022,Yang2022}, and GRB 211227A \citep{Lu2022}---have been associated with a claimed kilonova afterglow \citep{Zhu2022}. We note that, even with the very large disk masses produced in our models, the reservoir of mass is insufficient to sustain accretion and non-trivial jet luminosity for the long durations required to match these observations. However, accretion might be sustained by fallback \citep{Metzger2010Fallback}, as suggested in \cite{Metzger2010Fallback}, \cite{Desai2019}, and \cite{Zhu2022}. 
Alternatively if the magnetic field is configured to magnetically arrest the disk, accretion may be further sustained.

In this work, we have demonstrated that a large diversity of outcomes is possible from BH-NS merger remnants. To fully connect this disk modeling to observations, more models covering a denser sampling of parameter space, simulations run to later times, and full radiative transfer models of the kilonova are needed. These studies will be the subject of future work.

\section{Acknowledgements}

We thank Francois Foucart for discussions on BH-NS mergers and impact on the remnant, Luke Roberts for discussions on $r$-process nucleosynthesis, and Josh Dolence, Gail McLaughlan, Oleg Korobkin, and Chris Fryer for many useful interactions.
We also thank our anonymous referee for their constructive report, which significantly strengthened the manuscript.
Research presented in this work was supported by the Laboratory Directed Research and Development program of Los Alamos National Laboratory (LANL) through the Centers for Earth and Space Science (CSES) and Nonlinear Studies (CNLS) under project numbers 20210528CR, 20220545CR-CNL, and 20220564ECR. 
This research also used resources provided by LANL through the institutional computing program.
LANL is operated by Triad National Security, LLC, for the National Nuclear Security Administration of U.S. Department of Energy (Contract No. 89233218CNA000001).
The work at NC State was supported by United States Department of Energy (DOE), Office of Science, Office of  Nuclear Physics under Award DE-FG02-02ER41216. 
This work is approved for unlimited release with LA-UR-22-32395.

\software{
nubhlight, 
PRISM, 
Python, 
NumPy, SciPy, 
Matplotlib 
}

\bibliography{main_rev1}{}
\bibliographystyle{aasjournal}


\end{document}